\documentclass[twocolumn,prb,showpacs,preprintnumbers,superscriptaddress,amsmath,amssymb,citeautoscript]{revtex4-1}
\pdfoutput=1
\usepackage{graphicx}
\usepackage{epstopdf}
\usepackage{dcolumn}
\usepackage{color}
\usepackage{amsmath,amssymb}
\usepackage[english]{babel}
\usepackage{bm}
\usepackage[hyperindex]{hyperref}
\hypersetup{colorlinks,citecolor=blue, filecolor=blue,linkcolor=blue,urlcolor=blue}

\graphicspath{{./figures/}}
\usepackage{multirow}
\usepackage{rotating}
\usepackage{array}

\newcommand{\bpm}{\begin{pmatrix}}
\newcommand{\epm}{\end{pmatrix}}
\newcommand{\bs}{\boldsymbol}
\newcommand{\be}{\begin{equation}}
\newcommand{\ee}{\end{equation}}
\newcommand{\beq}{\begin{eqnarray}}
\newcommand{\eeq}{\end{eqnarray}}

\DeclareMathOperator{\im}{Im}

\DeclareMathOperator{\tr}{tr}
\DeclareMathOperator{\Ln}{Ln}

\DeclareMathOperator*{\res}{res}
\DeclareMathOperator{\ctg}{ctg}

\begin{document}

\title{Exact lattice-model calculation of boundary modes for Weyl semimetals and graphene}

\author{Vardan Kaladzhyan}
\email{vardank@kth.se}
\affiliation{Department of Physics, KTH Royal Institute of Technology, Stockholm, SE-106 91 Sweden}
\author{Sarah Pinon}
\affiliation{Institut de Physique Th\'eorique, Universit\'e Paris Saclay, CEA
CNRS, Orme des Merisiers, 91190 Gif-sur-Yvette Cedex, France}
\author{Jens H. Bardarson}
\affiliation{Department of Physics, KTH Royal Institute of Technology, Stockholm, SE-106 91 Sweden}
\author{Cristina Bena}
\affiliation{Institut de Physique Th\'eorique, Universit\'e Paris Saclay, CEA
CNRS, Orme des Merisiers, 91190 Gif-sur-Yvette Cedex, France}

\date{\today}

\begin{abstract}
We provide an exact analytical technique to obtain within a lattice model the wave functions of the edge states in zigzag- and bearded-edge graphene, as well as of the Fermi-arc surface states in Weyl semimetals described by a minimal bulk model. We model the corresponding boundaries as an infinite scalar potential localized on a line, and respectively within a plane. We use the T-matrix formalism to obtain the dispersion and the spatial distribution of the corresponding boundary modes. Furthermore, to demonstrate the power of our approach, we write down the surface Green's function of the considered Weyl semimetal model, and we calculate the quasiparticle interference patterns originating from an impurity localized at the respective surface.
\end{abstract}

\maketitle

\section{Introduction}

Systems which exhibit edge states, be they topological or not, one-, two- or three-dimensional, superconducting or normal, have come under intense scrutiny over the past years. Among such examples one can mention graphene\cite{Geim2007,deHeer2007,CastroNeto2009}, topological insulators and topological superconductors\cite{Kane2005,Qi2011,Bernevig2013}, Weyl semimetals\cite{Armitage2018}, and many others. Traditional techniques to calculate the wave functions associated with the boundary modes include numerical techniques such as tight-binding exact diagonalization\cite{Slater1954,Haydock1972,Haydock1975,Falicov1975,Busch1987,Kim2015}, solving the Schr\"odinger equation with the corresponding boundary conditions\cite{Berry1987,Davison1992,Li2010,Delplace2011,Zhang2012,Gorbar2015,Devizorova2017,Duncan2018}, using the bulk-boundary correspondence\cite{Hatsugai1993,Mong2011,Fukui2012,Kunst2017,Rhim2018} and the extended Bloch theorem\cite{Kunst2019}.

Here we focus on a qualitatively different approach to solving the problem of finding boundary modes, which was first introduced for Majorana bound states in Refs.~[\onlinecite{Zazunov2016,Zazunov2017,Kaladzhyan2019}], and subsequently developed in Refs.~[\onlinecite{Pinon2019a}] and [\onlinecite{Pinon2019b}] for topological insulators and Andreev bound states, respectively. In this work we extend the technique to derive analytically the boundary modes of Dirac systems. The core of the technique consists of modeling the boundary as a point-, line-, or plane-like localized scalar impurity potential with an infinite amplitude. This model can be solved exactly using the T-matrix formalism and allows to obtain in a very elegant and straightforward analytical manner the energy dispersion and the wave function of the bound states, as well as the surface/edge Green's functions of the system.

%Here we focus on a qualitatively different approach to solving the problem of finding boundary modes, which was introduced in Ref.~[\onlinecite{Kaladzhyan2019}]. The core of the technique consists of modeling the boundary as a point-, line-, or plane-like localized scalar impurity potential with an infinite amplitude. This model can be solved exactly using the T-matrix formalism and allows to obtain in a very elegant and straightforward analytical manner the energy dispersion and the wave function of the bound states, as well as the surface/edge Green's functions of the system.

%Here we focus on a qualitatively different approach to solving the problem of finding boundary modes, which was introduced in Ref.~[\onlinecite{Kaladzhyan2019}] for Majorana states and was successfully applied in Ref.~[\onlinecite{Pinon2019a}] to study the formation of boundary modes in topological insulators and Weyl semimetals, and of Andreev bound states in Ref.~[\onlinecite{Pinon2019b}]. The core of the technique consists of modeling the boundary as a point-, line-, or plane-like localized scalar impurity potential with an infinite amplitude. This model can be solved exactly using the T-matrix formalism and allows to obtain in a very elegant and straightforward analytical manner the energy dispersion and the wave function of the bound states, as well as the surface/edge Green's functions of the system.

In this paper we focus on deriving exact closed-form analytical expressions for the edge modes of zigzag- and bearded-edge graphene, as well as for their three-dimensional generalization --- Fermi-arc surface states in Weyl semimetals described by a minimal tight-binding model\cite{Turner2013}. Our results are consistent with previous findings obtained using different techniques, both in graphene and in Weyl semimetals, including, e.g., a recursive evaluation of the edge states in a tight-binding model\cite{Fujita1996,Wakabayashi2010a,Wakabayashi2010b,Bellec2014}, and analytical studies of the Schr\"odinger equation with specific boundary conditions in graphene\cite{Akhmerov2008,MessiasdeResende2017,Araujo2019} and Weyl semimetals \cite{Okugawa2014,Hashimoto2017}. Additionally, using the analytical result for the surface Green's functions, we calculate the quasiparticle interference patterns originating from a localized impurity at the surface of the considered Weyl semimetal.  We should stress that our method provides a number of advantages with respect to more traditional ones, in that it does not require any numerical algorithms, such as recursive Green's function calculation\cite{Sancho1985,Sancho1984,Umerski1997,Peng2017} or exact diagonalization, and therefore, can significantly speed up other calculations that require finding the boundary modes (e.g., in transport simulations). Moreover, in certain cases, like the ones described in this work, it yields exact closed-form expressions for the surface Green's functions, the edge states and the surface states, which provides one with more insight into the physics of the problem. Also, our technique is general, being applicable to any tight-binding model on any type of lattice structure in any number of dimensions. Finally, we are not required to make any low-energy approximations, as we can employ the full lattice model for the system under consideration. 

The paper is organized as follows: in Sec.~II we obtain the energy dispersion and the wave functions of the boundary modes for both zigzag- and bearded-edge graphene. We derive the Fermi-arc surface states and compute quasiparticle interference patterns in a Weyl semimetal in Sec.~III, and we leave the conclusions to Sec.~IV.

\section{Zigzag- and bearded-edge modes in graphene}
The simplest lattice model for graphene can be written as $H_0 = \int \frac{d\bs{k}}{(2\pi)^2} \Psi^\dag_{\bs{k}} \mathcal{H}_0(k_x,k_y)\Psi_{\bs{k}} $, with $\Psi_{\bs{k}} \equiv \{\psi^A_{\bs{k}},\, \psi^B_{\bs{k}} \}^\mathrm{T} $, where the indices $A$ and $B$ refer to the corresponding sublattices, and
\begin{align}\label{eq:HGraphene}
\nonumber &\mathcal{H}_{0}(k_x,k_y) = \\
  &t
	\bpm
		0 & e^{-i k_x} \negthickspace+\negthickspace 2 e^{i\frac{k_x}{2}} \cos \frac{k_y \sqrt{3}}{2} \\
		e^{i k_x} \negthickspace +\negthickspace 2 e^{-i\frac{k_x}{2}} \cos \frac{k_y \sqrt{3}}{2} & 0
	\epm,
\end{align}
where we set the lattice constant to unity and $t$ is the hopping amplitude. In what follows we express all energies in units of the hopping amplitude, equivalently we set  $t=1$. The first Brillouin zone for this model is defined as a hexagon with corners located at
$
(k_x,k_y)=\left( \pm \frac{2\pi}{3},\,\pm\frac{2\pi}{3\sqrt{3}} \right),\; 
\left(0,\,\pm\frac{4\pi}{3\sqrt{3}} \right).
$
%where every pair is written in the form $(k_x,\,k_y)$, and all the pairwise choices of signs can be made. 
The bare Matsubara Green's function for the Hamiltonian in Eq.~(\ref{eq:HGraphene}) can be calculated using the standard definition:
\begin{align}
\nonumber &G_0(k_x,k_y,i\omega_{\ell}) \equiv \left[i\omega_{\ell} - \mathcal{H}_0(k_x,k_y) \right]^{-1} = \\
\nonumber & -\frac{1}{3+2 \cos k_y \sqrt{3} + 4 \cos \frac{k_y \sqrt{3}}{2} \cos \frac{3k_x}{2} - (i\omega_{\ell})^2} \times \\
	&\bpm
		i\omega_{\ell} & e^{-i k_x} + 2 e^{i\frac{k_x}{2}} \cos \frac{k_y \sqrt{3}}{2} \\
		e^{i k_x} + 2 e^{-i\frac{k_x}{2}} \cos \frac{k_y \sqrt{3}}{2} & i\omega_{\ell}
	\epm.
\end{align}
where $\omega_{\ell} = \pi T \left(2 \ell + 1 \right)$, with $\ell \in \mathbb{Z}$, denote the fermionic Matsubara frequencies. 

\begin{figure}[ht!]
	\centering
	\includegraphics[width=0.9\columnwidth]{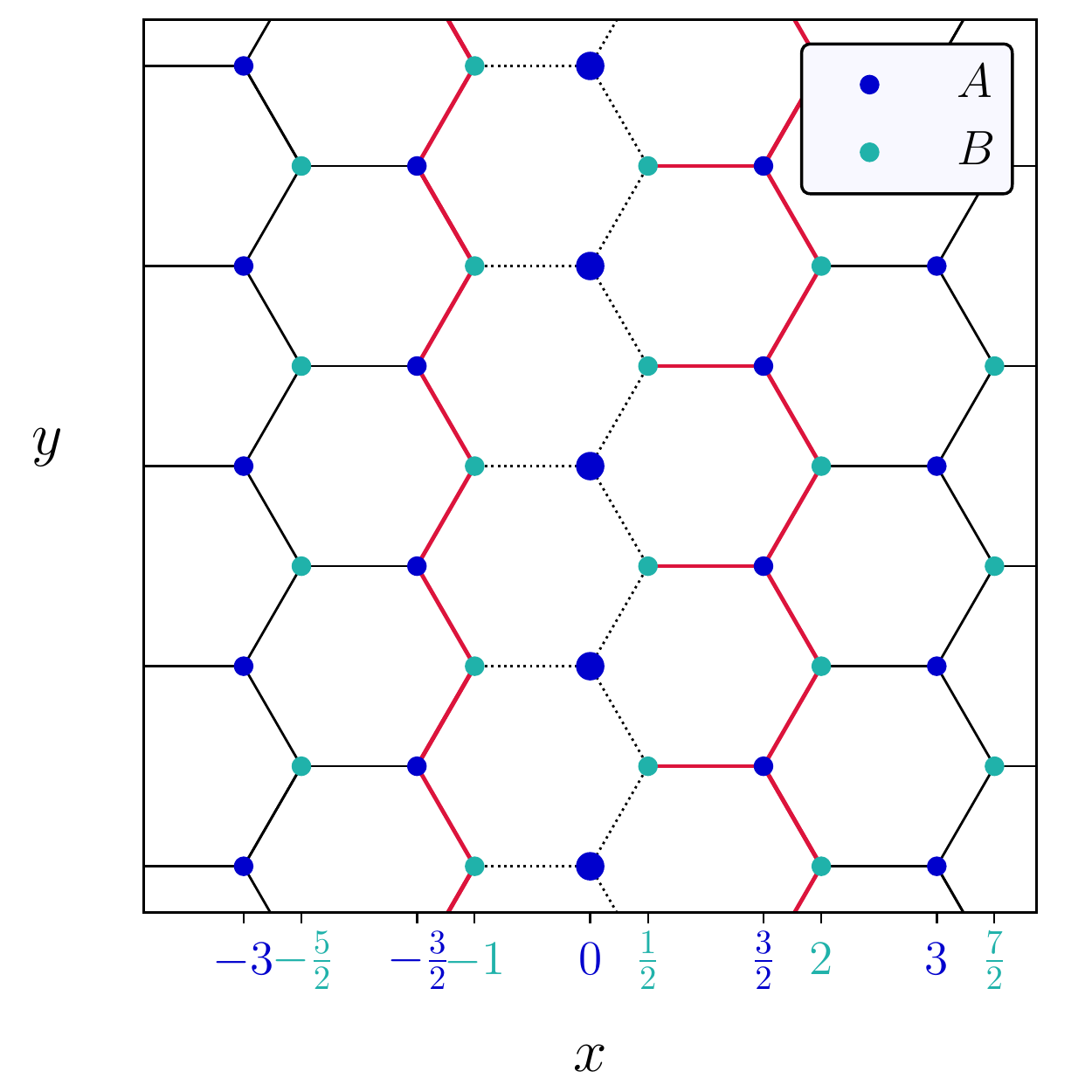}
	\caption{Graphene lattice with an infinite-amplitude $\delta$-function impurity introduced at $x=0$ (impurity sites are thus effectively disconnected from the lattice, as indicated by the dashed lines). Effectively we have a zigzag edge half-plane on the left side of the impurity, i.e., for $x<0$, and a bearded edge half-plane on the right side of the impurity, i.e., for $x>0$ (red lines).}
\label{fig:honeycomb}
\end{figure}

Our goal is to find the boundary modes for the two types of edges we are interested in: zigzag and bearded edges (shown in red in Fig.~\ref{fig:honeycomb}). In order to model an edge along the $y$ axis, we introduce an impurity potential localized solely on the atoms of sublattice A, as shown by large blue circles at $x=0$ in Fig.~\ref{fig:honeycomb}. Such an impurity can be described by $V = U \sum\limits_{n \in \mathbb{Z}} \left[\psi^{A}_{(0,n\sqrt{3})}\right]^\dag \psi^A_{(0,n\sqrt{3})} $, with $(0,n\sqrt{3})$ being the real-space lattice points. This choice of the impurity potential in the limit of $U \to \infty$ divides the infinite graphene sheet into two halves, at $x<0$ and $x>0$, where the former has a zigzag edge and the latter a bearded one, both terminating on atoms of sublattice B (see the green lattice sites in Fig.~\ref{fig:honeycomb}). 

To find the impurity-induced states, which evolve into boundary modes when $U$ is much larger than the other energy scales in the model, we calculate the bare Matsubara Green's function in the mixed coordinate space and momentum space representation, keeping in mind that the momentum along the $y$ direction $k_y$ is in this configuration a good quantum number:
\begin{align}
\label{eq:GFkxintegral}
G_0(x,k_y,i\omega_{\ell}) \equiv \int\limits_{-2\pi/3}^{2\pi/3} \frac{dk_x}{4\pi/3} G_0(k_x,k_y,&i\omega_{\ell}) e^{i k_x x}.
\end{align}
The integration limits $\pm 2\pi/3$ and the numerical factor $4\pi/3$ were derived in Ref.~[\onlinecite{Pinon2019a}] for a hexagonal lattice. Note also that in the expression above, $x$ can only take discrete values corresponding to the positions of the atoms in the honeycomb lattice.
These values can be divided into two classes, corresponding to the atoms of the sublattices A and B, 
\begin{align}\label{eq:XAXBpoints}
x_A = \frac{3n}{2}\;\text{and}\; x_B = \frac{3n+1}{2},\;\text{with}\; n \in \mathbb{Z}. 
\end{align}
We note that when $n < 0$ we are dealing with lattice sites on the left side of the impurity line, whereas for $n > 0$ with those on the right side. 

In order to perform the integration in Eq.~(\ref{eq:GFkxintegral}), we first denote:
\begin{align}
\label{eq:Xmdefinitions}
X_m(x,k_y,i \omega_{\ell}) \equiv -\negthickspace\int\limits_{-2\pi/3}^{2\pi/3}\negthickspace \frac{dk_x}{4\pi/3} \frac{e^{i k_x m/2} e^{i k_x x}}{D}, 
\end{align}
where $D \equiv 3+2 \cos k_y \sqrt{3} + 4 \cos \frac{k_y \sqrt{3}}{2} \cos \frac{3k_x}{2} - (i\omega_{\ell})^2$, $m \in \{0,\,\pm1,\,\pm2\}$. Due to the translational invariance in the $y$ direction, the integrals above are periodic functions of $k_y$, with a period given by $\left[0, \frac{2\pi}{\sqrt{3}} \right]$. The Green's function in Eq.~(\ref{eq:GFkxintegral}) can thus be written as
\begin{align}\label{eq:GFGraphene}
\nonumber &G_0(x,k_y,i\omega_{\ell}) = \\	
	&\bpm
		i\omega_{\ell} X_0& 2 X_1 \cos \frac{k_y \sqrt{3}}{2} + X_{-2} \\
		2 X_{-1} \cos \frac{k_y \sqrt{3}}{2} + X_{2} & i\omega_{\ell} X_0
	\epm,
\end{align}
where for simplicity we have omitted the explicit arguments of the $X_m$ functions. The detailed calculation of the five  $X_m$ integrals is presented in Appendix \ref{AppGraphene}. 

\begin{figure}[ht!]
	\centering
	\includegraphics[width=7cm,trim={0cm 0cm 0cm 0cm},clip]{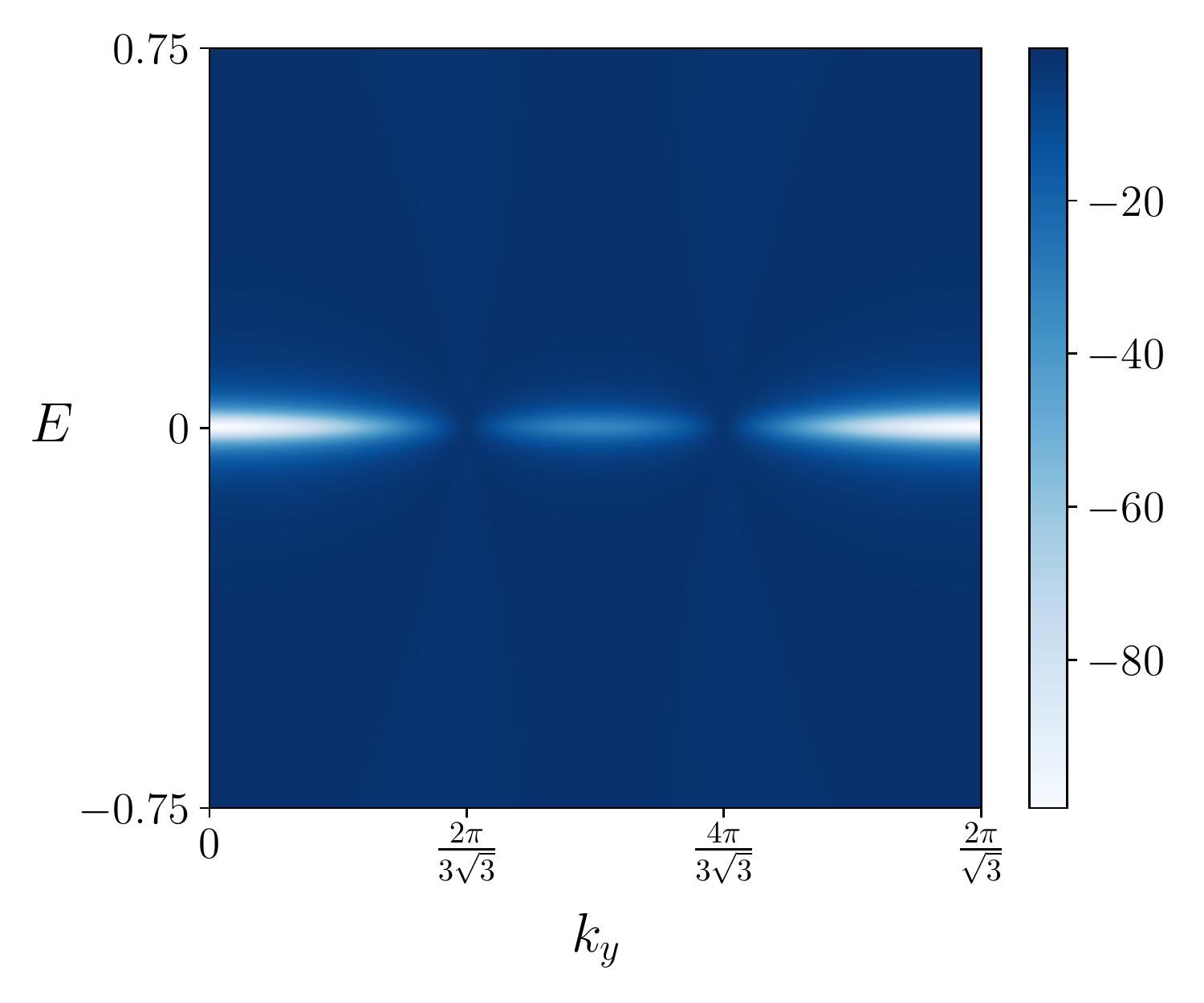}
	\caption{Imaginary part of the trace of the T-matrix, as computed in Eq.~(\ref{eq:GrapheneTmatrix}). At $E = 0$, we recover the edge states corresponding to both the zigzag and the bearded edges for $k_y \in \left(\frac{2\pi}{3\sqrt{3}}, \frac{4\pi}{3\sqrt{3}} \right)$ and $k_y \in \left[0, \frac{2\pi}{\sqrt{3}} \right]\backslash \left[\frac{2\pi}{3\sqrt{3}}, \frac{4\pi}{3\sqrt{3}} \right]$, respectively.}
\label{fig:GrapheneTmatrix}
\end{figure}

Below, using the Green's function in Eq.~(\ref{eq:GFGraphene}) we compute the $T$-matrix in order to find its poles defining the energies of the edge states of graphene in the limit where the impurity potential amplitude $U$ is the largest energy scale in the system. The $T$-matrix can be found as follows\cite{Mahan2000,Balatsky2006} $T \equiv \left(\mathbb{I} - V G_0 \right)^{-1} V$, which in our case becomes:
\begin{align}
\nonumber &T(k_y, i\omega_\ell) \\
\nonumber &\equiv \left[ \bpm 1& 0\\0 & 1\epm - \bpm U& 0\\0 & 0\epm \cdot G_0(x=0,k_y, i\omega_{\ell}) \right]^{-1} \cdot \bpm U& 0\\0 & 0\epm  \\
& \xrightarrow[U \to \infty]{} -\bpm \left[G_0^{11}(x=0,k_y, i\omega_{\ell}) \right]^{-1}& 0\\0 & 0\epm,
\end{align}
where $G_0^{11}$ denotes the $11$ component of the corresponding Green's function. We perform the analytical continuation $i\omega_\ell \to E + i\delta$, $\delta \to +0$, and we find that the imaginary part of the trace of the $T$-matrix,
\begin{align}\label{eq:GrapheneTmatrix}
\nonumber & \im \tr T(k_y, E + i0) = \\
&\im\frac{\sqrt{\left(3+2\cos k_y \sqrt{3} - (E + i0)^2\right)^2-16 \cos^2 \frac{k_y \sqrt{3}}{2}}}{E + i0} 
\end{align}
has a pole at $E=0$ for all $k_y \in \left[0, \frac{2\pi}{\sqrt{3}} \right] \backslash \left\{ \frac{2\pi}{3\sqrt{3}},\frac{4\pi}{3\sqrt{3}} \right\}$ (see Fig.~\ref{fig:GrapheneTmatrix}). 
Thus, there exists a zero-energy edge state for all the possible values of $k_y$ except for the special points $\left\{ \frac{2\pi}{3\sqrt{3}},\frac{4\pi}{3\sqrt{3}} \right\}$. It is known from literature\cite{Wakabayashi2010a,Wakabayashi2010b,Bellec2014} that if we had only one type of edge---bearded or zigzag---we would only recover edge modes for specific values of $k_y$. Namely, for the zigzag edge we would have a zero-energy edge state only for $k_y \in \left(\frac{2\pi}{3\sqrt{3}}, \frac{4\pi}{3\sqrt{3}} \right)$, whereas for the bearded edge we would have it for $k_y \in \left[0, \frac{2\pi}{\sqrt{3}} \right]\backslash \left[\frac{2\pi}{3\sqrt{3}}, \frac{4\pi}{3\sqrt{3}} \right]$. The special points $\frac{2\pi}{3\sqrt{3}}$ and $\frac{4\pi}{3\sqrt{3}}$ are the points where the zero-energy state cease to be edge states and merge with the bulk. For our particular configuration we have both zigzag and bearded edges, and thus it is not surprising that the existence of the edge states extends to the entire range of $k_y$. 

To find the wave functions corresponding to the edge states we use the same algorithm as for the Yu-Shiba-Rusinov states  \cite{Yu1965,Shiba1968,Rusinov1969,Sakurai1970,Pientka2013,Kaladzhyan2016}. First, we obtain the wave function at the impurity position ($x=0$) using
\begin{align}
\left[ \bpm 1& 0\\0 & 1\epm \negthickspace -\negthickspace G_0(x=0,k_y,E=0) \negthickspace\cdot\negthickspace \bpm U& 0\\0 & 0\epm \right] \Psi(x\negthickspace=\negthickspace 0, k_y) = 0.
\end{align}
We then use the propagation relation 
\begin{align}\Psi(x, k_y) = G_0(x,k_y,E=0) \cdot \bpm U& 0\\0 & 0\epm \Psi(x=0, k_y)
\end{align}
 to recover the dependence of the wave function on $x$, and we get:
\begin{align}\label{eq:edgestateWF}
\Psi(x, k_y) = \bpm 0 \\ X_{0}(x_B+1) + 2\cos\frac{k_y \sqrt{3}}{2} X_{0}(x_B-\frac{1}{2}) \epm,
\end{align}
where we have set $E = 0$ and we have omitted the $k_y$ dependence in $X_{0}$ for the sake of brevity. The explicit definitions of $x_B$ and $X_0$ are given in Eqs.~(\ref{eq:XAXBpoints}) and (\ref{eq:Xmdefinitions}), and the exact expression for the latter can be found in Appendix \ref{AppGraphene}. Note that this result is in qualitative agreement with previous studies of zigzag edge states \cite{Fujita1996,Nakada1996,Brey2006,Akhmerov2008,Yao2009,Wakabayashi2010a,Wakabayashi2010b,Bellec2014} in that the edge wave function is nonzero only on the atoms of the sublattice B. The wave function is thus defined only for $x = x_B \equiv \frac{3n+1}{2}$ (see Eq.~(\ref{eq:XAXBpoints})), and it is equal to zero for $x = x_A$.

\begin{figure}
	\centering
	\includegraphics[width=9cm,trim={0cm 0cm 0cm 0cm},clip]{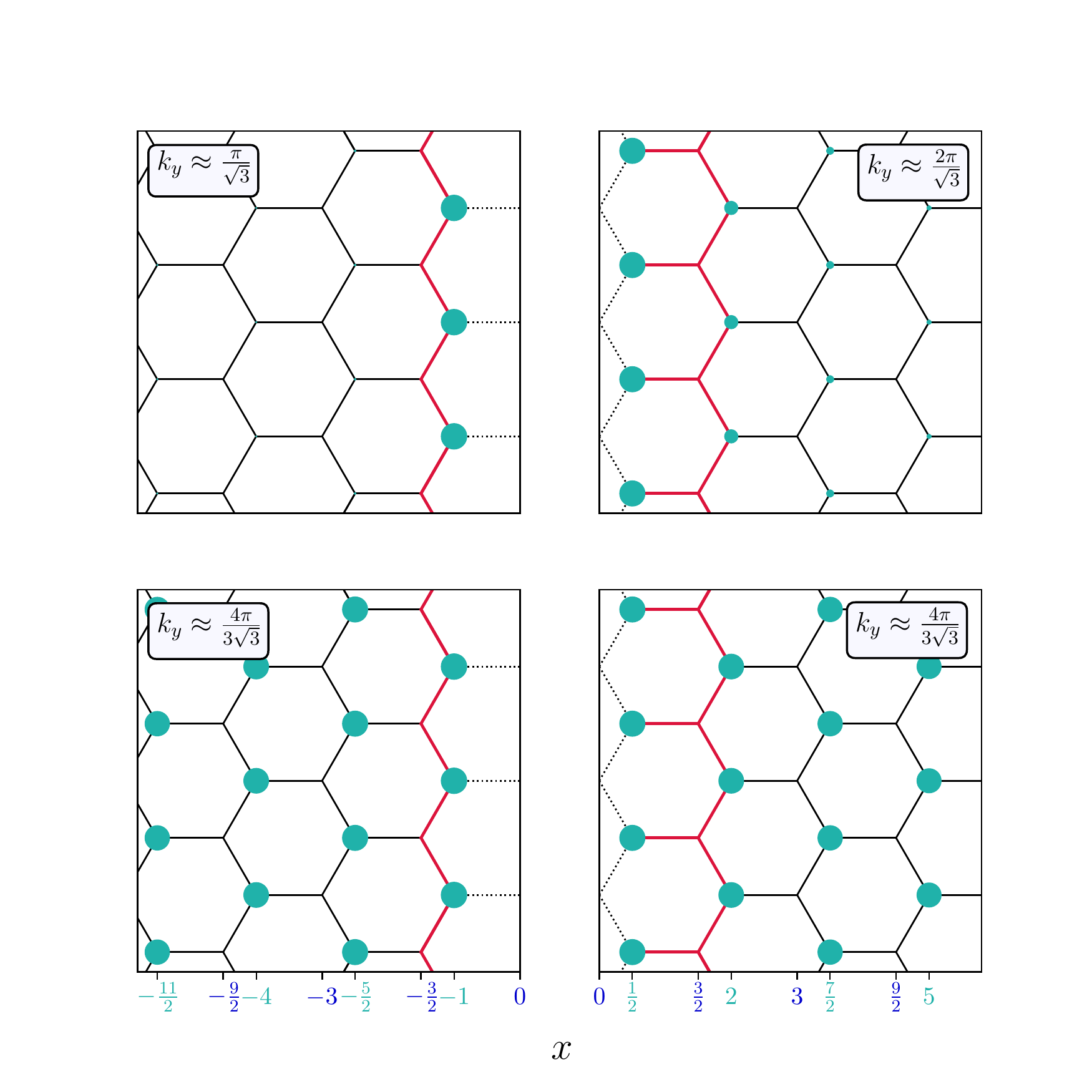}
	\caption{The LDOS calculated for the wave function in Eq.~(\ref{eq:edgestateWF}) for four different values of $k_y$. In the left and right columns we show the zigzag and bearded edges, respectively. The values of $k_y$ on the upper two panels correspond to the cases in which the edge states are the most localized, i.e., the localization length as a function of $k_y$ reaches a minimum, whereas the lower two panels show the opposite case, where the edge states are the most delocalized. The size of the circles reflects the magnitude of the LDOS.}
\label{fig:ZBWF}
\end{figure} 

Further simplifications can be made assuming that $k_y$ lies in one of the two previously mentioned intervals. The wave function in Eq.~(\ref{eq:edgestateWF}) then simplifies to: 
\begin{align}
\Psi(n,k_y) &= \bpm 0 \\ -\left(-2\cos\frac{k_y \sqrt{3}}{2}\right)^{-n-1} \epm,
\end{align}
which yields the zigzag-edge wave functions for $n < 0$, $k_y \in \left(\frac{2\pi}{3\sqrt{3}}, \frac{4\pi}{3\sqrt{3}} \right)$ and the bearded-edge wave functions for $n > 0$, $k_y \in \left[0, \frac{2\pi}{\sqrt{3}}\right] \backslash \left[\frac{2\pi}{3\sqrt{3}}, \frac{4\pi}{3\sqrt{3}} \right]$. Other combinations of $n$ and $k_y$ yield non-normalizable solutions, and thus can be discarded. Note the exact equivalence to previous results obtained using recursive calculations\cite{Fujita1996,Wakabayashi2010a,Wakabayashi2010b,Bellec2014} and specific boundary conditions\cite{Akhmerov2008,Delplace2011}.

In Fig.~\ref{fig:ZBWF} we plot the corresponding lattice local density of states (LDOS) $\rho(x,k_y) = |\Psi(x, k_y)|^2$. For values of $k_y$ between $\frac{2\pi}{3\sqrt{3}}$ and $\frac{4\pi}{3\sqrt{3}}$ the LDOS calculated from the wave function in Eq.~(\ref{eq:edgestateWF}) is nonzero only for $x<0$, corresponding to the zigzag edge state. Contrary to that, if the value of $k_y$ is chosen to be outside of the aforementioned range, the LDOS is nonzero only for $x>0$, and therefore corresponds to the bearded edge state.

\section{Weyl semimetal: Fermi arcs and Quasiparticle Interference Patterns}

In this section we turn to calculating the wave functions for the Fermi-arc states and the quasiparticle interference patterns on the surface of a Weyl semimetal described by the following model:
\begin{align}\label{eq:HWeyl}
\nonumber \mathcal{H}_{0}(\bs{k}) &= v\sin k_x \sigma_x + v \sin k_y \sigma_y + \\
&(m - t\cos k_x - t\cos k_y - t\cos k_z) \sigma_z.
\end{align}
The above Hamiltonian is defined on a cubic lattice with the lattice constant set to unity. Here $\bs{k} \equiv (k_x,\,k_y,\,k_z)$, the Pauli matrices are denoted by $(\sigma_x, \sigma_y, \sigma_z)$, $v$ characterizes the group velocity of the low-energy Weyl fermions, $m$ is the mass term, and $t$ is the hopping amplitude. In what follows we express all quantities with the dimensionality of energy in terms of the hopping amplitude, and thus we set $t=1$. We assume also for simplicity that $v = 1$. For $1 < m < 3$ the Hamiltonian above describes a Weyl semimetal phase. The Weyl nodes appear along the $k_z$ axis, at the two points defined by $\cos k^0_z = m - 2$. Note that this is a simple minimal model exhibiting only two Weyl nodes in the Brillouin zone. The bare Matsubara Green's function is defined as follows:
\begin{align}
\label{eq:WeylBulkGF}
 G_0(\bs{k},i\omega_{\ell}) &\equiv \left[ i\omega_{\ell} - \mathcal{H}_0(\bs{k}) \right]^{-1} = \\
\nonumber &-\frac{i\omega_\ell \sigma_0 + \tilde{g} \sigma_z + \sin k_x \sigma_x + \sin k_y \sigma_y}{\tilde{g}^2 + \sin^2 k_x + \sin^2 k_y -(i\omega_\ell)^2},
\end{align}
where we denote $\tilde{g} \equiv \tilde{g}(\bs{k}) \equiv m  - \cos k_x - \cos k_y - \cos k_z$.
%, and $\omega_{\ell} = \pi T \left(2 \ell + 1 \right)$.

\subsection{Fermi-arc surface states}
In order to find the Fermi-arc states we introduce into the system  a plane-like boundary at $y=0$, emulated by a $\delta$-function impurity with the following potential $V(y) = \bpm U& 0\\0 & U\epm\delta(y)$. To solve the corresponding impurity problem we compute the Matsubara Green's function in the mixed coordinate space and momentum space representation, i.e.,
\begin{align}
\label{eq:GFkyintegral}
G_0(k_x,y,k_z,i\omega_{\ell}) \equiv \int\limits_{-\pi}^{\pi} \frac{dk_y}{2\pi} G_0(\bs{k},&i\omega_{\ell}) e^{i k_y y}.
\end{align}
In the expression above $y$ is not a continuous variable since the system is defined on a lattice, therefore, it admits only integer values (both positive and negative, depending on whether the system lies at $y \geqslant 0$ or $y \leqslant 0$), in other words, $y = n a$, where $a=1$ is the lattice constant, and $n \in \mathbb{Z}$. To perform the Fourier transform in  Eq.~(\ref{eq:GFkyintegral}) we define three integrals:
\begin{align}\label{eq:Xsdefintions}
X_s &\equiv -\negthickspace\int\limits_{-\pi}^{\pi}\negthickspace \frac{dk_y}{2\pi} \frac{ (\bullet)\, e^{i k_y y}}{\tilde{g}^2 + \sin^2 k_x + \sin^2 k_y -(i\omega_\ell)^2},
\end{align}
where $(\bullet) = 1,\, \cos k_y$ and $\sin k_y$ for $s=0,\, 1$ and $2$, respectively. We leave the step-by-step calculations of the integrals above to Appendix \ref{AppWSM}. In terms of these integrals the Green's function can be written as
\begin{align}\label{eq:GFWeyl}
\nonumber G_0(k_x,y,k_z,i\omega_{\ell}) = \phantom{aaaaaaaaaaaaaaaaaaaaaaaaa}\\	
i\omega_{\ell} X_0 \sigma_0 + (g X_0 - X_1) \sigma_z + \sin k_x X_0 \sigma_x + X_2 \sigma_y,
\end{align}
where we denote $g \equiv g(k_x,k_z) \equiv m  - \cos k_x - \cos k_y$. Here we have omitted the arguments of the $X_s$ functions for the sake of brevity. 

In what follows we compute the $T$-matrix using the mixed real and reciprocal space representation of the bare Green's function given in Eq.~(\ref{eq:GFWeyl}), and assuming that the impurity potential amplitude $U$ is the largest energy scale in the system, i.e., $U \gg 1$. Thus, we get
\begin{align}\label{eq:WeylTmatrix}
\nonumber T(k_x, k_z, i\omega_\ell) &\equiv \left[ \bpm 1& 0\\0 & 1\epm -U G_0(k_x,y=0,k_z, i\omega_{\ell}) \right]^{-1} \\
&\xrightarrow[U \to \infty]{} - \left[ G_0(k_x,y=0,k_z, i\omega_{\ell}) \right]^{-1}.
\end{align}
The poles of the imaginary part of the trace of the $T$-matrix found via analytical continuation $i\omega_\ell \to E+i\delta, \delta \to +0$ define the energies of the impurity-bound states at finite values of $U$, whereas at $U \to \infty$ they yield the surface modes, i.e., the Fermi-arc surface states. We perform the analytical continuation, and we find that 
\begin{align}\label{eq:WeylTmatrixtr}
\nonumber \tr T(k_x, k_z, E+i0) = \phantom{aaaaaaaaaaaaaaaaaaa}\\
\frac{2 (E+i0) X_0}{\left(g X_0 + X_1\right)^2 + \left[ \sin^2k_x - (E+i0)^2 \right] X_0^2 }
\end{align}
whose imaginary part has two poles, at $E = \pm \sin k_x$, for $k_z$ lying in the interval between the Weyl nodes, i.e., $\left(-k_z^0, k_z^0 \right)$, and $k_x$ such that $\cos k_x \geqslant m - 1 - \cos k_z$. Outside of these regions, the $T$-matrix does not have any poles, and therefore, the Fermi arcs cease to exist. It is worth noting that there are two solutions, $E = +\sin k_x$ and $E = - \sin k_x$, due to the fact that a plane-like impurity introduces both a `top' surface and a `bottom' surface, one for the Weyl semimetal in the lower half-space, $y \leqslant 0$, and another one for that in the upper half-space, $y \geqslant 0$.

\begin{figure}
	\centering
	\includegraphics[width=8cm,trim={0cm 0cm 0cm 0cm},clip]{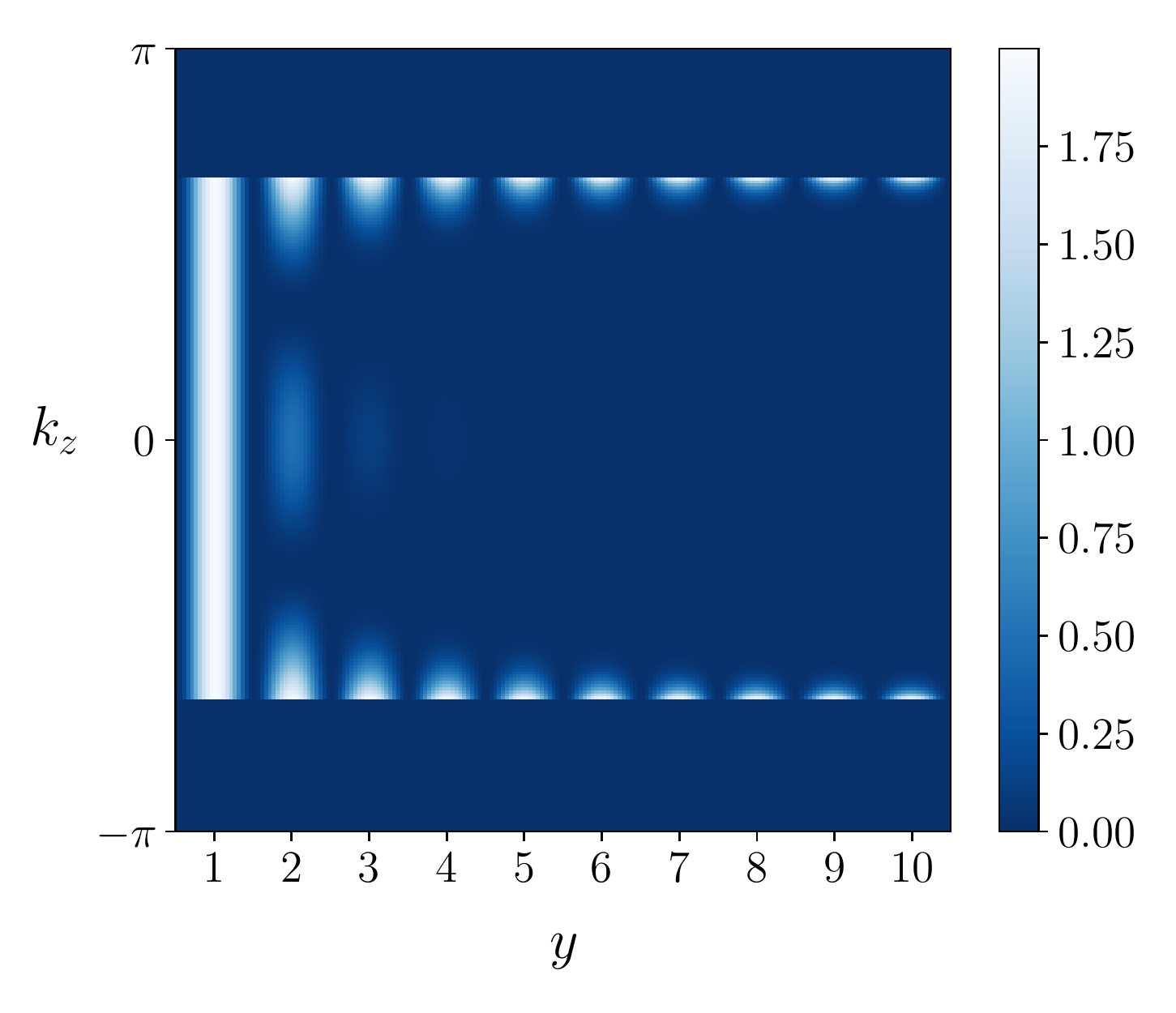}
	\caption{The LDOS for the solution at $E = 0$, as defined by Eq.~(\ref{eq:LDOSWeyl}), plotted as a function of $y$ and $k_z$. The mass term is set to $m = 1.5$, yielding Weyl nodes at $k^0_z = \pm 2\pi/3$, in the vicinity of which the Fermi-arc surface state becomes more delocalized, and the corresponding LDOS  becomes more bulk-like.}
\label{fig:FAWF}
\end{figure}

To study the spatial dependence of the Fermi-arc wavefunctions we use the same procedure as in the previous section \cite{Yu1965,Shiba1968,Rusinov1969,Sakurai1970,Pientka2013,Kaladzhyan2016}. First, we find the wave function at $y=0$ from the equation: 
$$
G_0(k_x,y=0,k_z, E = \pm\sin k_x) \Psi_{\pm}(k_x,y=0,k_z) = 0,
$$ 
whose non-trivial solutions can be found by imposing $\det G_0(k_x,y=0,k_z, E = \pm\sin k_x) =0 $, which in turn yields:
\begin{align}
E &= +\sin k_x \to \Psi_+(k_x,y=0,k_z) \propto \bpm 1\\ -1 \epm,\\
E &= -\sin k_x \to \Psi_-(k_x,y=0,k_z) \propto \bpm 1\\ 1 \epm.
\end{align}
The solution for $E = +\sin k_x$ corresponds to the Weyl semimetal lying in the upper half-space, whereas the one for $E = +\sin k_x$ to that in the lower half-space. The $y$-dependence of the wave function can be found using $\Psi(k_x,y,k_z) = G_0(k_x,y,k_z, E) \cdot U \Psi(k_x,y=0,k_z)$:
\begin{align}
\Psi_+(k_x,y,k_z) &= (g X_0 - X_1 + i X_2) \bpm 1 \\ 1 \epm \\ 
\Psi_-(k_x,y,k_z) &= (g X_0 - X_1 - i X_2) \bpm 1 \\ -1 \epm.
\end{align}
The physics encoded in these two wave functions is qualitatively the same, and thus in Fig.~(\ref{fig:FAWF}) we plot the LDOS for one of the solutions at $E = 0$ (equivalently, $k_x = 0$), i.e., 
\begin{equation}
\label{eq:LDOSWeyl}
\rho(k_x=0,y,k_z,E = 0) \equiv |\Psi_+(k_x=0,y,k_z)|^2
\end{equation} 
as a function of $y \geqslant 0$, at different values of $k_z$. It is clear that the localization length of the Fermi-arc surface state changes with $k_z$, and as expected, the state becomes completely delocalized exactly at the Weyl nodes, i.e., at $k_z^0 = \pm 2\pi/3$. 

\subsection{Quasiparticle interference pattern} 

To demonstrate the  power of our approach for solving boundary problems, in what follows we calculate the quasiparticle interference patterns associated to the presence of a single localized impurity $V_s(x,z) = \bpm U_s & 0 \\ 0 & U_s \epm \delta(x,z)$ on the surface of a Weyl semimetal. Along the lines of Ref.~[\onlinecite{Pinon2019a}], the surface Green's function for the Weyl semimetal in the upper half plane ($y>0$) is given by:

\begin{align}
\label{eq:WeylSurfaceGF}
G_s(k_x,k_z) = \negthickspace\int\negthickspace\frac{dk_{1y}dk_{2y}}{(2\pi)^2} G(k_x, k_{1y}, k_{2y}, k_z) e^{i (k_{1y}-k_{2y}) y} \Big|_{y=1},
\end{align}
with
\begin{align}
\label{eq:WeylBulkFullGF}
\nonumber & G(k_x, k_{1y}, k_{2y}, k_z) = 2\pi G_0(k_x, k_{1y}, k_z) \delta(k_{1y}-k_{2y}) + \\
& G_0(k_x, k_{1y}, k_z) T(k_x, k_z) G_0(k_x, k_{2y}, k_z),
\end{align}
where  the unperturbed bulk Green's function and the $T$-matrix are given by Eqs.~(\ref{eq:WeylBulkGF}) and (\ref{eq:WeylTmatrix}), respectively. We set $y = 1$ in the expression above because in the presence of an infinite-amplitude $\delta$-potential impurity all sites at $y = 0$ are cut out of the system, thus moving the surface of the system to $y = 1$.
Above we omitted writing down the explicit energy dependence in all Green's functions. We rewrite the integrand in Eq.~(\ref{eq:WeylSurfaceGF}) using Eq.~(\ref{eq:WeylBulkFullGF}), and we obtain the surface Green's function performing both integrals over momenta in Eq.~(\ref{eq:WeylSurfaceGF}): 
\begin{align}
\label{eq:WeylSurfaceGFfinal}
\nonumber & G_s(k_x,k_z) = G_0(k_x, y=0, k_z) + \\
& G_0(k_x, y = 1, k_z) T(k_x, k_z) G_0(k_x, y = -1, k_z),
\end{align}
where the mixed coordinate space-momentum space representation of the unperturbed Green's function is given by Eq.~(\ref{eq:GFWeyl}).  

\begin{figure}[t!]
	\centering
	\includegraphics[width=0.9\columnwidth]{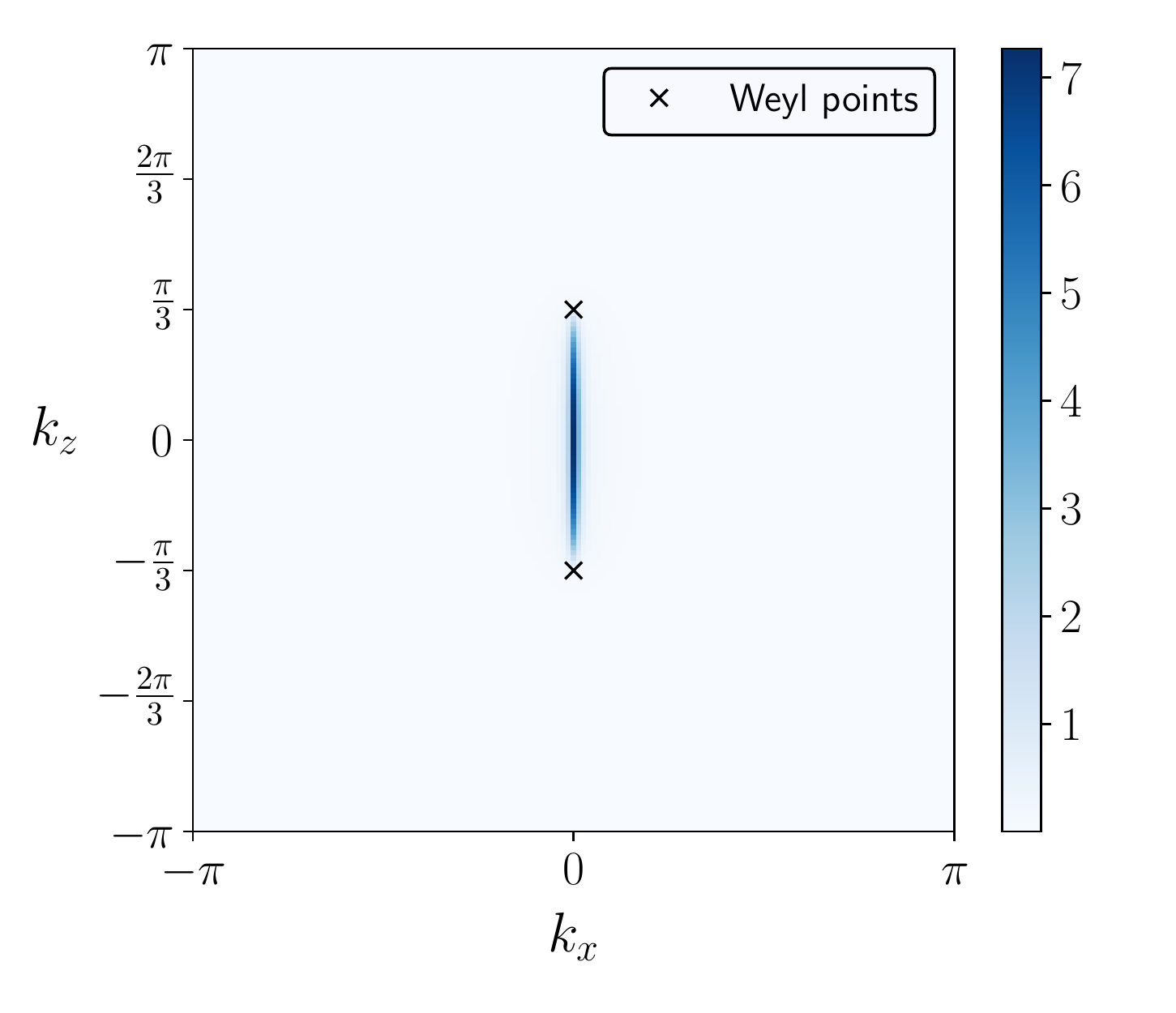}
	\caption{Spectral function at $E = 0$, as defined by Eq.~(\ref{eq:WeylSurfaceSpectralFunction}), plotted as a function of $k_x$ and $k_z$. We set $m = 2.5$, thus Weyl nodes appear at $k^0_z = \pm \pi/3$. We can clearly see the Fermi arc (in this toy model, a line) connecting the two nodes.}
\label{fig:SpectralFunction}
\end{figure}

In what follows, we study the spectral function defined through the surface Green's function in Eq.~(\ref{eq:WeylSurfaceGFfinal}) as follows:
\begin{align}
\label{eq:WeylSurfaceSpectralFunction}
A(k_x,k_z) \equiv -\frac{1}{\pi} \im \tr G_s(k_x,k_z).
\end{align}
In Fig.~\ref{fig:SpectralFunction} we plot the spectral function taken at $E=0$, and we observe a line in the momentum space, corresponding to the Fermi arc. Since we employ one of the simplest lattice models of a Weyl semimetal (i.e., obtained by stacking Chern insulators in reciprocal space), the Fermi-arc surface states at a fixed value of energy appear as lines in the momentum space. Therefore, \textit{a priori} we expect a very simple quasiparticle interference pattern in this toy-model case, namely, the scattering at the surface occurs mostly between the surface states, along with some residual scattering into the bulk as well.

Below we define the quasiparticle interference patterns in the momentum space via
\begin{align}
\label{eq:QPIMS}
\Delta \rho (k_x, k_z) = -\frac{1}{2\pi i}\negthickspace\int\negthickspace\frac{dq_x dq_z}{(2\pi)^2} f(q_x,q_z,k_x,k_z)
\end{align}
where
\begin{align*}
f \equiv \tr \big[ & G_s(q_x,q_z) T_s G_s(q_x+k_x,q_z+k_z) - \\
& G^*_s(q_x+k_x,q_z+k_z) T^*_s G^*_s(q_x,q_z)\big].
\end{align*}
Also
\begin{align*}
T_s = \left[\bpm 1 & 0 \\ 0 & 1\epm - U_s \int \frac{dk_x dk_z}{(2\pi)^2} G_s(k_x,k_z) \right]^{-1} U_s,
\end{align*}
 $U_s$ is the amplitude of the impurity potential, and $*$ denotes complex conjugation.

\begin{figure}[t!]
	\centering
	\includegraphics[width=0.9\columnwidth]{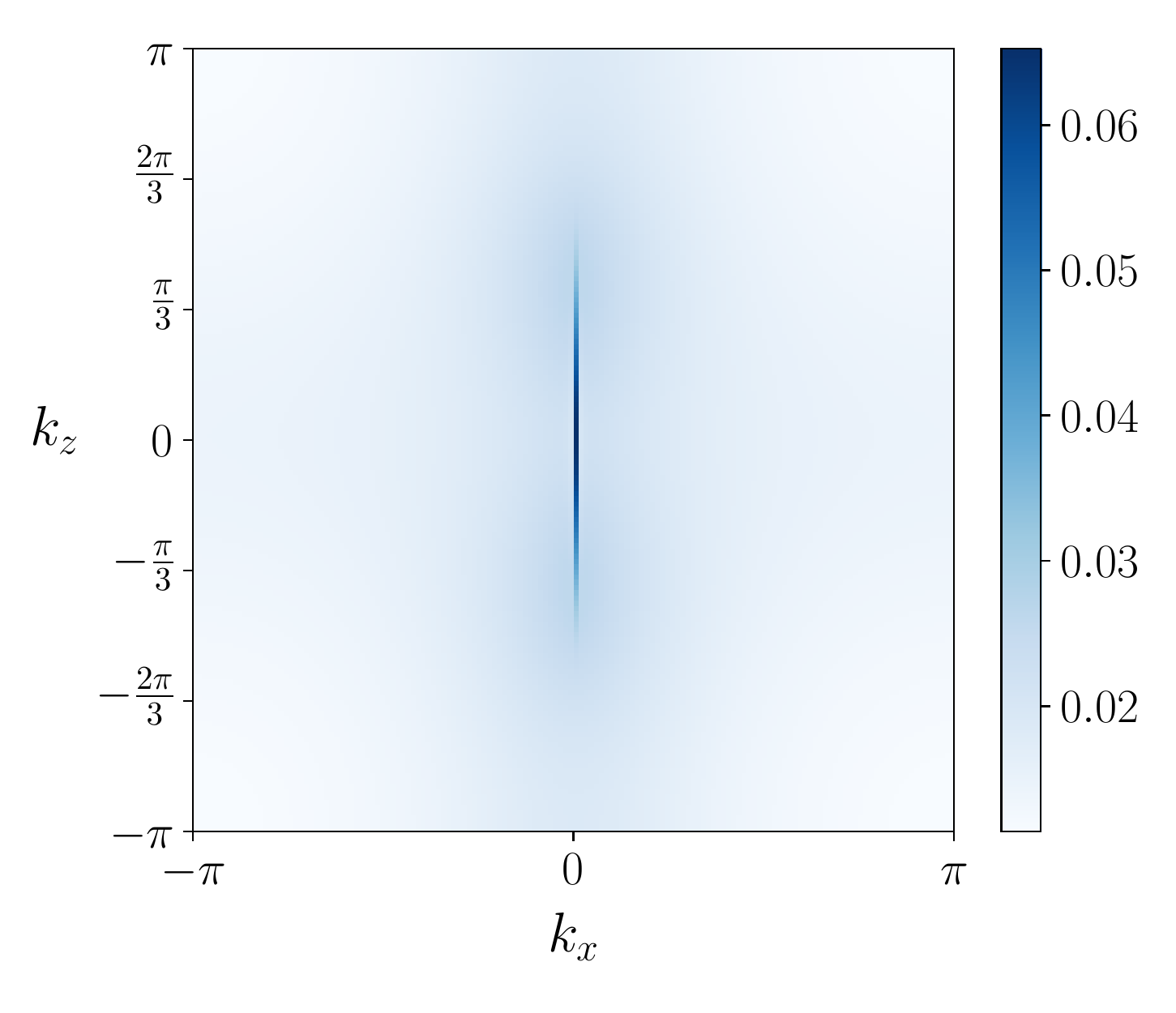}
	\caption{Quasiparticle interference pattern at $E = 0$, as defined by Eq.~(\ref{eq:QPIMS}), plotted as a function of $k_x$ and $k_z$. The mass term is set to $m = 2.5$, yielding Weyl nodes at $k^0_z = \pm \pi/3$. The impurity potential amplitude is set to $U_s=1$. The strong line-like feature in the middle reflects the surface-surface scattering processes, while the small background ($\Delta \rho \approx 10^{-2}$) accounts for the surface-bulk and bulk-bulk ones.}
\label{fig:QPIMS}
\end{figure}

In Fig.~\ref{fig:QPIMS} we plot the quasiparticle interference pattern for the case of a shorter Fermi arc, i.e., when the Weyl nodes lie within the interval $\left[-\pi/2, \pi/2\right]$, or in other words, when $k_z^0 < \pi/2$. Most of the weight in the plot is concentrated, as expected, in a narrow line lying from $-2k_z^0$ to $-2k_z^0$ at $k_x = 0$, which corresponds to surface-surface scattering processes. Indeed, most of the scattering occurs within the Fermi arc, and the maximum change in momentum along the $k_z$ axis is $2k_z^0$, while along the $k_x$ axis it is $0$. The very small and undifferentiated background corresponds to surface-bulk and bulk-bulk scattering processes. In the case of a longer Fermi arc (e.g., $m = 1.5$), the overall change in $k_z$ can go beyond the first Brillouin zone, and thus the narrow line is expected to lie in the interval $\left[-\pi, \pi \right]$. 

More complex and realistic models for Weyl semimetals exhibiting topological Fermi-arc states were treated in Ref.~[\onlinecite{Pinon2020}], with a stark focus on spin-resolved components of quasiparticle interference patterns and the surface Green's function derivation employing the technique from Refs.~[\onlinecite{Zazunov2016,Zazunov2017,Kaladzhyan2019,Pinon2019a,Pinon2019b}]. Most importantly, no closed-form analytical solutions for the wave functions of the Fermi-arc states are presented in Ref.~[\onlinecite{Pinon2020}], contrary to the present work which mostly focuses on analytical derivations of boundary-mode wave functions.

\section{Conclusions}
To summarize, we have proposed an analytical route to calculate the boundary modes of graphene and Weyl semimetals within a lattice model. For the Weyl semimetals we have considered a minimal tight-binding model exhibiting two cones and a line-like Fermi arc, and we have also computed the quasiparticle interference patterns via a calculation of the surface Green's function. Our results are obtained by modeling  the boundaries as localized infinite-amplitude impurity potentials and treating the problem exactly within the T-matrix formalism. More specifically, we have recovered an exact closed form for the wave functions of the zigzag and the bearded edge states for graphene, as well as for the Fermi-arc surface states and surface Green's function for a Weyl semimetal model considered. The results presented in this work are in agreement with previous calculations performed either by recursive tight-binding methods\cite{Fujita1996,Wakabayashi2010a,Wakabayashi2010b}, or by solving the Schr\"odinger equation with specifically derived boundary conditions\cite{Delplace2011,Hashimoto2017}, or other methods\cite{Kunst2017,Kunst2019}.

The technique we have employed is very general and can be applied to any lattice model no matter its complexity or its dimensionality, and it is in no way limited to only minimal tight-binding models, but can be used for \textit{any} tight-binding model. It allows to recover energies of the boundary modes, as well as exact forms for their wave functions without requiring any numerical calculations such as, e.g., exact diagonalization, except at most a numerical integral of the Green's function to Fourier transform it from momentum space to real space; however, in the present work we have obtained all integrals in a closed analytical form.

Potential applications include, but are not limited to: using the technique in setups with pseudomagnetic fields\cite{Peri2020}; studying disordered systems numerically by treating a single disorder realization via the generalized T-matrix or analytically by introducing a boundary-emulating impurity after averaging over disorder; considering more realistic models for studying quasiparticle interference patterns in Weyl and Dirac semimetals\cite{Pinon2020}.

\begin{acknowledgments}
V.K. and J.H.B. would like to acknowledge the ERC Starting Grant No. 679722. V.K. acknowledges also the Roland Gustafsson foundation for theoretical physics, and the Karl Engvers foundation, as well as Mark O. Goerbig, Cl\'ement Dutreix and Lo\"ic Herviou for fruitful discussions. 
\end{acknowledgments}

\bibliography{biblio_Tmatrix_GrapheneWeyl}

\newpage
\widetext
\appendix

\section{Derivation of the edge states for zigzag- and bearded-edge graphene}\label{AppGraphene}

In this Appendix we calculate the integrals in Eq.~(\ref{eq:Xmdefinitions}) for $m \in \{-2, -1, 0, 1, 2 \}$:
\begin{align}
\label{eq:AppXmdefinitions}
X_m(x,k_y,i \omega_{\ell}) \equiv -\negthickspace\int\limits_{-2\pi/3}^{2\pi/3}\negthickspace \frac{dk_x}{4\pi/3} \frac{e^{i k_x m/2} e^{i k_x x}}{3+2 \cos k_y \sqrt{3} + 4 \cos \frac{k_y \sqrt{3}}{2} \cos \frac{3k_x}{2} - (i\omega_{\ell})^2}, 
\end{align}
Translational invariance in the $y$ direction implies that the integral above is a periodic function of $k_y$, and we choose one period to be $k_y \in \left[0, \frac{2\pi}{\sqrt{3}} \right]$. The case of $k_y = \frac{\pi}{\sqrt{3}}$ should be considered separately since the integral simplifies significantly due to the fact that the denominator does not depend on $k_x$ anymore. Note also that the integral above is defined solely on the graphene lattice shown in Fig.~\ref{fig:honeycomb}, and thus $x = x_A = \frac{3n}{2}$ or $x = x_B = \frac{3n+1}{2}$, with $n \in \mathbb{Z}$.\\

We start with the case of $k_y = \frac{\pi}{\sqrt{3}}$, where we have:
\begin{align}
X_m(x,k_y,i \omega_{\ell}) = - \frac{1}{1 - (i\omega_{\ell})^2}\negthickspace\int\limits_{-2\pi/3}^{2\pi/3}\negthickspace \frac{dk_x}{4\pi/3} e^{i k_x m/2} e^{i k_x x} = -\frac{1}{1 - (i\omega_{\ell})^2} \frac{\sin \pi \left(\frac{2}{3}x + \frac{m}{3} \right)}{\pi \left(\frac{2}{3}x + \frac{m}{3} \right)}
\end{align}
When additionally $x=-m/2$, we get:
$
X_0\left( 0,\frac{\pi}{\sqrt{3}},i \omega_{\ell} \right) = -\frac{1}{1 - (i\omega_{\ell})^2}.
$\\

In what follows we turn to the case of $k_y \neq \frac{\pi}{\sqrt{3}}$. In this case the integral in Eq.~(\ref{eq:AppXmdefinitions}) can be rewritten as a contour integral in the complex plane, using the substitution $z = e^{\frac{3}{2}k_x}$:
\begin{align}
\label{eq:X_m_complexplane}
X_m(x,k_y,i \omega_{\ell}) \equiv -\frac{1}{2 \cos \frac{k_y \sqrt{3}}{2}}\; \frac{1}{2\pi i}\negthickspace \oint\limits_{|z|=1}\negthickspace dz \frac{z^{\frac{2}{3}\left(x+\frac{m}{2} \right)}}{z^2+fz+1},
\end{align}
where the circle $|z|=1$ is oriented counter-clockwise, and we defined
\begin{align}
f \equiv f(k_y, i\omega_{\ell}) = \frac{3+2 \cos k_y \sqrt{3} - (i\omega_{\ell})^2}{2 \cos \frac{k_y \sqrt{3}}{2}}.
\end{align}
For simplicity we rewrite this integral for $x$ lying on sublattices A and B separately:
\begin{align}
\label{eq:X^A_m_complexplane}&\text{sublattice A:}\quad X^A_m(x = \frac{\phantom{+}3n\phantom{1}}{2},k_y,i \omega_{\ell})\;\, = -\frac{1}{2 \cos \frac{k_y \sqrt{3}}{2}}\; \frac{1}{2\pi i}\negthickspace \oint\limits_{|z|=1}\negthickspace dz \frac{z^{n+m/3}}{z^2+fz+1}, \\
\label{eq:X^B_m_complexplane}&\text{sublattice B:}\quad X^B_m(x = \frac{3n+1}{2},k_y,i \omega_{\ell}) = -\frac{1}{2 \cos \frac{k_y \sqrt{3}}{2}}\; \frac{1}{2\pi i}\negthickspace \oint\limits_{|z|=1}\negthickspace dz \frac{z^{n+(m+1)/3}}{z^2+fz+1},
\end{align}
Before we proceed with the calculation, several important simplifications are worth pointing out:\\

\noindent $\bullet$ It is easy to notice that for sublattice A we have 
$
X^A_{-2}(x) = X^A_2(-x),\;  X^A_{-1}(x) = X^A_1(-x),
$
whereas for sublattice B $X^B_{-2}(x) = X^A_1(\frac{1}{2}-x)$,  $X^B_{-1}(x) = X^A_0(\frac{1}{2}-x)$, $X^B_0(x) = X^A_1(x-\frac{1}{2})$, $X^B_1(x) = X^A_2(x-\frac{1}{2})$, and $X^B_2(x) = X^A_0(x+1)$. Above we omitted the dependence on $k_y$ and $i\omega_\ell$. Therefore, it is sufficient to compute only the three integrals in Eq.~(\ref{eq:X^A_m_complexplane}), for $m=0$, $m=1$ and $m=2$, respectively, and for $x = x_A$ only, since all integrals on sublattice B can be expressed in terms of those on sublattice A.\\

\noindent $\bullet$ From the definition in Eq.~(\ref{eq:X^A_m_complexplane}) it is clear that the case of $m=0$ and the case of $m \neq 0$ (i.e., $m=1$ and $m=2$) should be considered separately. The reason for it is that when $m=0$ there are no functions in the integrand requiring to make a branch cut in the complex plane, thus significantly simplifying the integration.\\

\noindent $\bullet$ Another important remark can be made about the roots of the quadratic polynomial in the denominator of the integrand in Eq.~(\ref{eq:X^A_m_complexplane}). A simple analysis of the roots given by
\begin{align}\label{eq:Appzpmroots}
z_\pm = \frac{1}{2} \left[-f \pm \sqrt{f^2-4} \right]
\end{align}
shows that, first, both roots are always real for the values of the parameters in consideration. Second, one of the roots always belongs to the interval $(-1,1)$, namely,
\begin{align}
&\text{if}\; 0 \leqslant k_y < \frac{\pi}{\sqrt{3}},\quad\;\text{then}\; z_+ \in (-1,0),\; z_- < -1, \\
&\text{if}\; \frac{\pi}{\sqrt{3}} < k_y \leqslant \frac{2\pi}{\sqrt{3}},\;\text{then}\; z_- \in (0,+1),\; z_+ > +1. 
\end{align}
This means that regardless of the value of $k_y$ one of the poles of the integrand always lies inside the unit circle in the complex plane, and has to be taken into account while applying the residue theorem. Additionally, Vieta's formula dictates $z_+ z_- = 1$.\\

\subsection{The sub-case of \boldmath{$m=0$}}
\noindent Using the roots in Eq.~(\ref{eq:Appzpmroots}) we rewrite the integral in  Eq.~(\ref{eq:X^A_m_complexplane}) in the following form (omitting the prefactor):
\begin{align}
\frac{1}{2\pi i}\negthickspace \oint\limits_{|z|=1}\negthickspace dz \frac{z^{n}}{(z-z_+)(z-z_-)} = *
\end{align}
It is easy to prove that this integral is unchanged if one changes $n$ to $-n$, therefore, here we compute it only for $n>0$. In that case one of the poles lies inside the unit circle, whereas the other one -- outside. We start with the case in which $z_+$ lies inside the circle ($0 \leqslant k_y < \pi/\sqrt{3}$), and using the residue theorem, we get:
\begin{align}
 * = \res_{z=z_+} \frac{z^{n}}{(z-z_+)(z-z_-)} = - \frac{z_+^{|n|}}{z_- - z_+}
\end{align}
To get the expression for the case where $z = z_-$ is inside the unit circle, and $z = z_+$ is outside ($\pi/\sqrt{3} < k_y \leqslant 2\pi/\sqrt{3}$), we just need to exchange $z_+ \to z_-$ and vice versa in expression above. Taking into account the symmetry with respect to flipping the sign of $n$, we get the final expression:
\begin{align}
X^A_0\left(x = \frac{3n}{2},k_y,i \omega_{\ell}\right) = -\frac{1}{2 \cos \frac{k_y \sqrt{3}}{2}} 
\begin{cases}
	-\frac{z_+^{|n|}}{z_- - z_+}, &0 \leqslant k_y < \frac{\pi}{\sqrt{3}}
 \\
	+\frac{z_-^{|n|}}{z_- - z_+}, &\frac{\pi}{\sqrt{3}} < k_y \leqslant \frac{2\pi}{\sqrt{3}}
\end{cases}.
\end{align}

\subsection{The sub-cases of \boldmath{$m=1$} and \boldmath{$m=2$}}
\begin{figure}
	\centering
	\includegraphics[width=0.8\columnwidth]{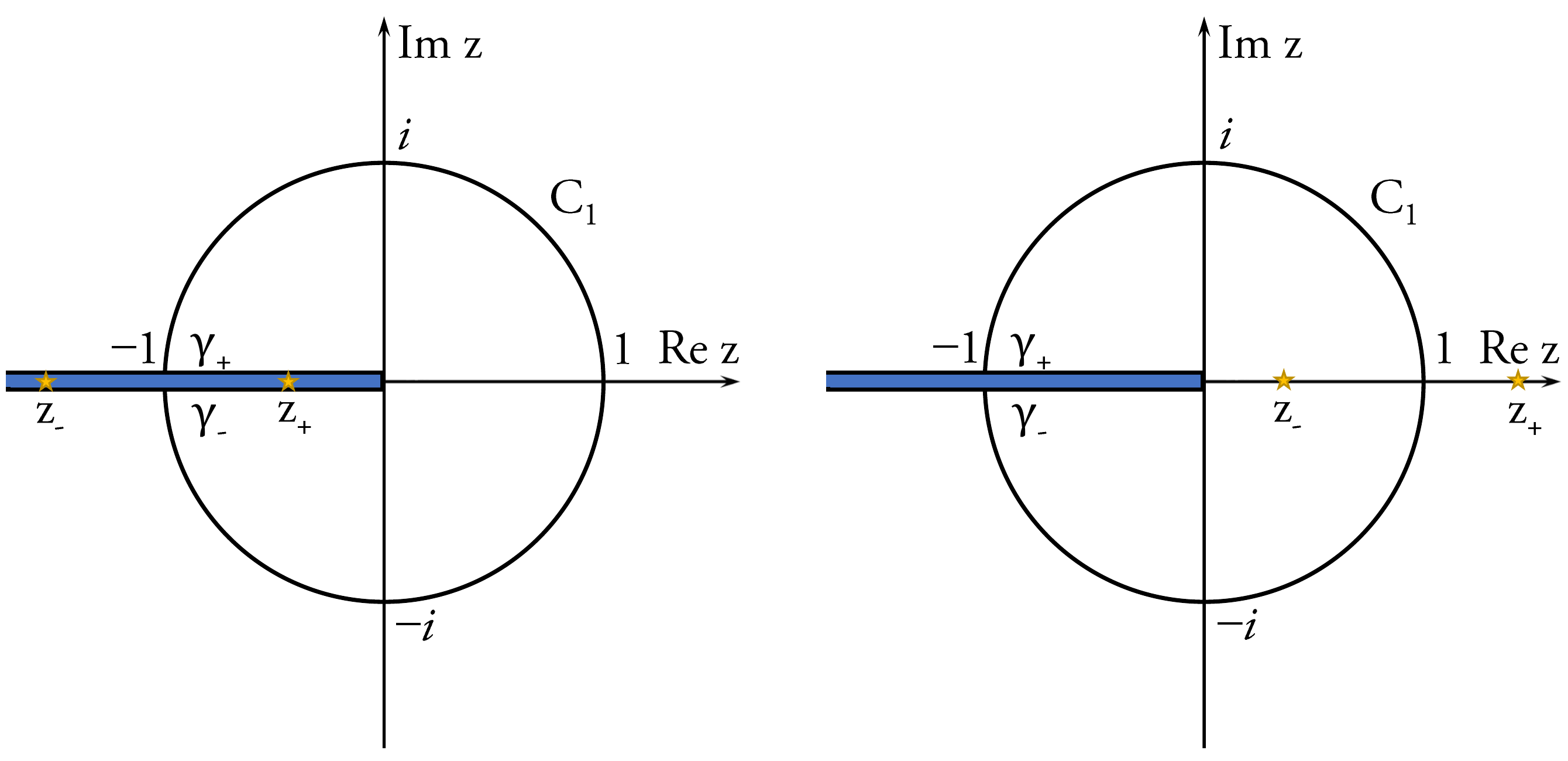}
	\label{fig:GammaBranchCutPole}
	\caption{Auxiliary contour $\Gamma$ for complex plane integration for $0 \leqslant k_y < \frac{\pi}{\sqrt{3}}$ (left panel) and $\frac{\pi}{\sqrt{3}} < |k_y| \leqslant \frac{2\pi}{\sqrt{3}}$ (right panel).}
\end{figure}

\noindent In this subsection we calculate the integral
\begin{align}
\frac{1}{2\pi i}\negthickspace \oint\limits_{|z|=1}\negthickspace dz \frac{z^{n+m/3}}{(z-z_+)(z-z_-)} = *
\end{align}
for $m=1$ or $m=2$. In this case we have to introduce a branch cut on the non-positive part of the real axis, i.e., for $z \in \left( -\infty, 0 \right]$. In order to start the calculation, we first define auxiliary contours in the complex plane, $\Gamma = C_1 \cup \gamma_+ \cup \gamma_-$ (see left and right panels of Fig.~\ref{fig:GammaBranchCutPole}). The unit circle is denoted $C_1$, and $\gamma_\pm$ are the right and left banks of the branch cut, parametrized by $t \pm i0$ with $t \in \left[ -1,0\right]$, correspondingly. Note, that to be entirely rigorous we should have added also a circle of infinitesimal radius around the point $z=0$, to ensure that we treat correctly the divergence at that point when $n<-1$, however, it appears that the method of calculation we apply takes care of that problem automatically.  For $0 \leqslant k_y < \pi/\sqrt{3}$ and $\pi/\sqrt{3} < k_y \leqslant 2\pi/\sqrt{3}$ we choose the left and the right panels of Fig.~\ref{fig:GammaBranchCutPole}, respectively. The difference between these panels is the position of the pole inside the unit circle: on the left panel, it falls into the branch cut, whereas on the right panel it lies within $(0,1)$. We note that in the definition of the auxiliary contour $\Gamma$ we neglect the tiny line between $\gamma_+$ and $\gamma_-$, i.e., $\left[-i0,+i0\right]$, since it never contributes to the value of the integral.\\

We start by considering the case of $0 \leqslant k_y < \pi/\sqrt{3}$ and we write down the residue theorem for the contour $\Gamma = C_1 \cup \gamma_+ \cup \gamma_-$ defined in the left panel of Fig.~\ref{fig:GammaBranchCutPole}. Since there are no poles inside the contour $\Gamma$ we have:
\begin{align}
\frac{1}{2\pi i}\negthickspace \oint\limits_{\Gamma} dz \frac{z^{n+m/3}}{(z-z_+)(z-z_-)} = 0 = \frac{1}{2\pi i} \left[\;\int\limits_{C_1} + \int\limits_{\gamma_+} + \int\limits_{\gamma_-} \right]\negthickspace dz \frac{z^{n+m/3}}{(z-z_+)(z-z_-)}. 
\end{align}
Therefore, we can express the sought-for line integral along $C_1 \equiv \{ \forall z: |z| = 1 \}$ as
\begin{align*}
* = -\frac{1}{2\pi i} \left[\;\int\limits_{\gamma_+} + \int\limits_{\gamma_-} \right]\negthickspace dz \frac{z^{n+m/3}}{(z-z_+)(z-z_-)} = **
\end{align*}
In order to compute the integrals along $\gamma_+$ and $\gamma_-$ we use a parametrisation $z = t+i0$, $t \in \left[ -1, 0 \right]$ and $z = t-i0$, $t \in \left[ 0, -1 \right]$, respectively. We start with $\gamma_+$ and we get:
\begin{align*}
\int\limits_{\gamma_+} \negthickspace dz \frac{z^{n+m/3}}{(z-z_+)(z-z_-)} &= \int\limits_{-1}^0 dt \frac{(t+i0)^{n+m/3}}{(t+i0-z_+)(t-z_-)} = \int\limits_{-1}^0 dt \frac{e^{(n+m/3)\Ln(t+i0)}}{(t+i0-z_+)(t-z_-)} = \int\limits_{-1}^0 dt \frac{e^{(n+m/3)\left[\ln|t| + \pi i \right]}}{(t+i0-z_+)(t-z_-)} = \\
&= e^{i\pi(n+m/3)}\int\limits_{-1}^0 dt \frac{|t|^{n+m/3}}{(t+i0-z_+)(t-z_-)} = e^{i\pi(n+m/3)}\int\limits_{0}^1 dw \frac{w^{n+m/3}}{(w+z_+-i0)(w+z_-)} = \\
&= e^{i\pi(n+m/3)} \left[ \mathcal{P}\negthickspace \int\limits_{0}^1 dw \frac{w^{n+m/3}}{(w+z_+)(w+z_-)} +i\pi \int\limits_{0}^1 dw \frac{w^{n+m/3}}{w+z_-} \delta(w+z_+) \right]
\end{align*}
Similarly, we get:
\begin{align*}
\int\limits_{\gamma_-} \negthickspace dz \frac{z^{n+m/3}}{(z-z_+)(z-z_-)} = -e^{-i\pi(n+m/3)} \left[ \mathcal{P}\negthickspace \int\limits_{0}^1 dw \frac{w^{n+m/3}}{(w+z_+)(w+z_-)} -i\pi \int\limits_{0}^1 dw \frac{w^{n+m/3}}{w+z_-} \delta(w+z_+) \right]
\end{align*}
Therefore, combining two integrals from left and right banks, we obtain:
\begin{align*}
** = \frac{(-1)^{n+1}}{\pi}\left(I_\mathcal{P} \sin \frac{\pi m}{3} + \pi I_\delta \cos \frac{\pi m}{3} \right) = ***,
\end{align*}
where we defined
\begin{align}
I_\mathcal{P} &= \mathcal{P}\negthickspace \int\limits_{0}^1 dw \frac{w^{n+m/3}}{(w+z_+)(w+z_-)} = \frac{(-z_-)^{n+\frac{m}{3}} B\left[-\frac{1}{z_-},n+\frac{m}{3}+1,0 \right]-(-z_+)^{n+\frac{m}{3}} \left( B\left[-z_+,-n-\frac{m}{3},0 \right] + \pi \ctg \frac{\pi m}{3} \right)}{z_--z_+} \\
I_\delta &= \int\limits_{0}^1 dw \frac{w^{n+m/3}}{w+z_-} \delta(w+z_+) = \frac{(-z_+)^{n+m/3}}{z_--z_+}
\end{align}
Substituting $I_\mathcal{P}$ and $I_\delta$ into the equation above, we obtain after simplifications:
\begin{align*}
*** =  \frac{(-1)^{n+1}}{\pi} \sin \frac{\pi m}{3} \frac{(-z_-)^{n+\frac{m}{3}} B\left[-\frac{1}{z_-},n+\frac{m}{3}+1,0 \right]-(-z_+)^{n+\frac{m}{3}} B\left[-z_+,-n-\frac{m}{3},0 \right]}{z_--z_+}.
\end{align*}
Thus we get:
\begin{align*}
\frac{1}{2\pi i}\negthickspace \oint\limits_{|z|=1}\negthickspace dz \frac{z^{n+m/3}}{(z-z_+)(z-z_-)} = \frac{(-1)^{n+1}}{\pi} \sin \frac{\pi m}{3} \frac{(-z_-)^{n+\frac{m}{3}} B\left[-\frac{1}{z_-},n+\frac{m}{3}+1,0 \right]-(-z_+)^{n+\frac{m}{3}} B\left[-z_+,-n-\frac{m}{3},0 \right]}{z_--z_+},
\end{align*}
where $B[z,\alpha,\beta]$ is the incomplete Beta-function defined as:
\begin{align}
B\left[z,\alpha,\beta\right] \equiv \int\limits_0^z t^{\alpha-1}(1-t)^{\beta-1}dt.
\end{align}\\

In what follows we consider the remaining case of $\frac{\pi}{\sqrt{3}} < k_y \leqslant \frac{2\pi}{\sqrt{3}}$. First, we rewrite the integral as follows:
\begin{align}
\frac{1}{2\pi i}\negthickspace \oint\limits_{C_1}\negthickspace dz \frac{z^{n+m/3}}{(z-z_+)(z-z_-)} =  \frac{1}{z_--z_+} \left[ \frac{1}{2\pi i}\negthickspace \oint\limits_{C_1}\negthickspace dz \frac{z^{n+m/3}}{z-z_-} - \frac{1}{2\pi i}\negthickspace \oint\limits_{C_1}\negthickspace dz \frac{z^{n+m/3}}{z-z_+}\right]
\end{align}
The first integral in the sum can be expressed in terms of the second integral (with a parameter change) by means of a variable change $z = 1/w$. Note that such a variable change inverts the orientation of the integration contour, thus multiplying the result by $-1$:
\begin{align}
\frac{1}{2\pi i}\negthickspace \oint\limits_{C_1}\negthickspace dz \frac{z^{n+m/3}}{z-z_-} = \frac{1}{2\pi i}\negthickspace \oint\limits_{C_1}\negthickspace \frac{dw}{w^2} \frac{w^{-n-m/3}}{1/w-z_-} = -\frac{1}{z_-}\frac{1}{2\pi i} \negthickspace \oint\limits_{C_1}\negthickspace dw \frac{w^{-n-m/3-1}}{w-1/z_-} = -\frac{1}{z_-}\frac{1}{2\pi i} \negthickspace \oint\limits_{C_1}\negthickspace dw \frac{w^{-n-m/3-1}}{w-z_+}.
\end{align}
From the expression above we see that to obtain the first integral from the second one we need to replace $n \to -n-2$, $m \to 3-m$ and multiply the result by $-1/z_-$. Next, we compute the second integral
\begin{align}
\frac{1}{2\pi i}\negthickspace \oint\limits_{C_1}\negthickspace dz \frac{z^{n+m/3}}{z-z_+} 
\end{align}
by writing down the residue theorem for the contour $\Gamma = C_1 \cup \gamma_+ \cup \gamma_-$ defined in the panel panel of Fig.~\ref{fig:GammaBranchCutPole} for the second integral. Since there are no poles inside the contour $\Gamma$ we have:
\begin{align}
\frac{1}{2\pi i}\negthickspace \oint\limits_{\Gamma}\negthickspace dz \frac{z^{n+m/3}}{z-z_+} = 0 = \frac{1}{2\pi i} \left[\;\int\limits_{C_1} + \int\limits_{\gamma_+} + \int\limits_{\gamma_-} \right]\negthickspace dz \frac{z^{n+m/3}}{z-z_+}. 
\end{align}
Therefore, we can express the sought-for line integral along $C_1 \equiv \{ \forall z: |z| = 1 \}$ as
\begin{align*}
\frac{1}{2\pi i}\negthickspace \oint\limits_{C_1}\negthickspace dz \frac{z^{n+m/3}}{z-z_+}  = - \frac{1}{2\pi i} \left[\;\int\limits_{\gamma_+} + \int\limits_{\gamma_-} \right]\negthickspace dz \frac{z^{n+m/3}}{z-z_+} = *
\end{align*}
In order to compute the integrals along $\gamma_+$ and $\gamma_-$ we use a parametrisation $z = t+i0$, $t \in \left[ -1, 0 \right]$ and $z = t-i0$, $t \in \left[ 0, -1 \right]$, respectively:
\begin{align}
\int\limits_{\gamma_\pm} \negthickspace dz \frac{z^{n+m/3}}{z-z_+} = \pm \int\limits_{-1}^0 dt \frac{(t\pm i0)^{n+m/3}}{t-z_+} = \pm e^{\pm i\pi(n+m/3)}\int\limits_{0}^1 dw \frac{w^{n+m/3}}{w+z_+}
\end{align}
Thus, we get: 
\begin{align*}
*= - \frac{1}{2\pi i} \left[\;\int\limits_{\gamma_+} + \int\limits_{\gamma_-} \right]\negthickspace dz \frac{z^{n+m/3}}{z-z_+} = - \frac{(-1)^n}{2\pi i} \left[e^{i\frac{\pi m}{3}}-e^{-i\frac{\pi m}{3}} \right]\int\limits_{0}^1 dw \frac{w^{n+m/3}}{w+z_+} = -\frac{1}{\pi} \sin \frac{\pi m}{3} e^{-i\frac{\pi m}{3}} z_+^{n+\frac{m}{3}} B\left[-\frac{1}{z_+},n+\frac{m}{3}+1,0 \right]
\end{align*}
Using the parameter substitution introduced above, we get the first integral:
\begin{align}
\frac{1}{2\pi i}\negthickspace \oint\limits_{C_1}\negthickspace dz \frac{z^{n+m/3}}{z-z_-}  = \frac{1}{\pi} \sin \frac{\pi (3-m)}{3} e^{-i\frac{\pi (3-m)}{3}} z_-^{n+\frac{m}{3}} B\left[-z_-,-n-\frac{m}{3},0 \right] = -\frac{1}{\pi} \sin \frac{\pi m}{3} e^{i\frac{\pi m}{3}} z_-^{n+\frac{m}{3}} B\left[-z_-,-n-\frac{m}{3},0 \right]
\end{align}
Finally:
\begin{align}
\frac{1}{2\pi i}\negthickspace \oint\limits_{|z|=1}\negthickspace dz \frac{z^{n+m/3}}{(z-z_+)(z-z_-)} = \frac{1}{\pi} \sin \frac{\pi m}{3}\frac{ e^{-i\frac{\pi m}{3}} z_+^{n+\frac{m}{3}} B\left[-\frac{1}{z_+},n+\frac{m}{3}+1,0 \right] - e^{i\frac{\pi m}{3}} z_-^{n+\frac{m}{3}} B\left[-z_-,-n-\frac{m}{3},0 \right]}{z_--z_+} 
\end{align}\\

Combining the results for different ranges of $k_y$, we present the final result:
\begin{align*}
X^A_m(x = \frac{3n}{2},k_y,i \omega_{\ell}) = -\frac{1}{2 \cos \frac{k_y \sqrt{3}}{2}} 
\begin{cases}
	\frac{(-1)^{n+1}}{\pi}\sin\frac{\pi m}{3} \frac{(-z_-)^{n+\frac{m}{3}} B\left[-\frac{1}{z_-},\,n+\frac{m}{3}+1,\,0 \right] - (-z_+)^{n+\frac{m}{3}} B\left[-z_+,\,-n-\frac{m}{3},\,0 \right] }{z_- - z_+} & 0 \leqslant k_y < \frac{\pi}{\sqrt{3}} \\
		\frac{1}{\pi} \sin \frac{\pi m}{3}\frac{ e^{-i\frac{\pi m}{3}} z_+^{n+\frac{m}{3}} B\left[-\frac{1}{z_+},n+\frac{m}{3}+1,0 \right] - e^{i\frac{\pi m}{3}} z_-^{n+\frac{m}{3}} B\left[-z_-,-n-\frac{m}{3},0 \right]}{z_--z_+} & \frac{\pi}{\sqrt{3}} < k_y \leqslant \frac{2\pi}{\sqrt{3}}
\end{cases},
\end{align*}
where $m=1$ or $m=2$.

\section{Derivation of the Fermi-arc states for a Weyl semimetal}\label{AppWSM}
In this Appendix we calculate the three integrals defined in Eq.~(\ref{eq:Xsdefintions}):
\begin{align}\label{eq:AppXsdefintions}
X_s &\equiv -\negthickspace\int\limits_{-\pi}^{\pi}\negthickspace \frac{dk_y}{2\pi} \frac{ (\bullet)\, e^{i k_y y}}{\tilde{g}^2 + \sin^2 k_x + \sin^2 k_y -(i\omega_\ell)^2},
\end{align}
where $(\bullet) = 1,\, \cos k_y$ and $\sin k_y$ for $s=0,\, 1$ and $2$, respectively. First, it is easy to see that integrals $X_1$ and $X_2$ can be expressed in terms of the integral $X_0$ in the following way:
\begin{align}
\nonumber X_1(k_x,y,k_z,i\omega_{\ell}) &= -\negthickspace\int\limits_{-\pi}^{\pi}\negthickspace \frac{dk_y}{2\pi} \frac{\cos k_y\, e^{i k_y y}}{D} = \frac{1}{2} \left[-\negthickspace\int\limits_{-\pi}^{\pi}\negthickspace \frac{dk_y}{2\pi} \frac{e^{i k_y (y+1)}}{D} -\negthickspace\int\limits_{-\pi}^{\pi}\negthickspace \frac{dk_y}{2\pi} \frac{e^{i k_y (y-1)}}{D} \right] = \\
&= \frac{1}{2}\left[X_0(k_x,y+1,k_z,i\omega_{\ell}) + X_0(k_x,y-1,k_z,i\omega_{\ell}) \right] \\
\nonumber X_2(k_x,y,k_z,i\omega_{\ell}) &= -\negthickspace\int\limits_{-\pi}^{\pi}\negthickspace \frac{dk_y}{2\pi} \frac{\sin k_y\, e^{i k_y y}}{D} = \frac{1}{2i} \left[-\negthickspace\int\limits_{-\pi}^{\pi}\negthickspace \frac{dk_y}{2\pi} \frac{e^{i k_y (y+1)}}{D} +\negthickspace\int\limits_{-\pi}^{\pi}\negthickspace \frac{dk_y}{2\pi} \frac{e^{i k_y (y-1)}}{D} \right] = \\
&= \frac{1}{2i}\left[X_0(k_x,y+1,k_z,i\omega_{\ell}) - X_0(k_x,y-1,k_z,i\omega_{\ell}) \right],
\end{align}
where for the sake of brevity we denoted $D \equiv \tilde{g}^2 + \sin^2 k_x + \sin^2 k_y -(i\omega_\ell)^2, \tilde{g} = m - \cos k_x - \cos k_y \cos k_z$. 
Therefore, we need to compute only the integral $X_0$:
\begin{align*}
X_0(k_x,y,k_z,i\omega_{\ell}) = -\negthickspace\int\limits_{-\pi}^{\pi}\negthickspace \frac{dk_y}{2\pi} \frac{e^{i k_y y}}{m^2+\omega_\ell^2 + (\cos k_x +\cos  k_y +\cos k_z)(-2m + \cos k_x +\cos  k_y +\cos k_z) + \sin^2 k_x + \sin^2 k_y} = \\
= -\negthickspace\int\limits_{-\pi}^{\pi}\negthickspace \frac{dk_y}{2\pi} \frac{e^{i k_y y}}{2(-m+\cos k_x + \cos k_z)\cos k_y + 1 + m^2 + \omega_\ell^2 + (\cos k_x + \cos k_z)(-2m + \cos k_x + \cos k_z) +\sin^2k_x} = *
\end{align*}
If $-m+\cos k_x + \cos k_z = 0$, then we have:
\begin{align}
* = -\frac{1}{1 + \sin^2k_x - (i\omega_\ell)^2} \negthickspace\int\limits_{-\pi}^{\pi}\negthickspace \frac{dk_y}{2\pi} e^{i k_y y} = -\frac{1}{1 +\sin^2k_x - (i\omega_\ell)^2} \delta_{y,0} 
\end{align}
If $-m+\cos k_x + \cos k_z \neq 0$, then
\begin{align}
\nonumber *&= \frac{1}{2(m - \cos k_x - \cos k_z)} \negthickspace\int\limits_{-\pi}^{\pi}\negthickspace \frac{dk_y}{2\pi} \frac{e^{i k_y y}}{\cos k_y - f(k_x,k_z)} = \frac{1}{2(m - \cos k_x - \cos k_z)} \frac{1}{2\pi i}\negthickspace\oint\limits_{|z|=1}\negthickspace \frac{dz}{z} \frac{z^y}{\frac{1}{2}(z+z^{-1}) - f(k_x,k_z)} = \\
\nonumber &= \frac{1}{m - \cos k_x - \cos k_z} \frac{1}{2\pi i}\negthickspace\oint\limits_{|z|=1}\negthickspace dz \frac{z^y}{z^2 - 2 f(k_x,k_z) z + 1} = **
\end{align}
Above we introduced
$$
f(k_x, k_z) \equiv \frac{1 + g^2 +\sin^2k_x - (i\omega_\ell)^2}{2g},\quad g(k_x,k_z) \equiv m - \cos k_x - \cos k_z.
$$
Note, that since $m \in (1,3)$, $g(k_x,k_z) \in (-1,1)$ for all values of $k_x$ and $k_z$. In what follows we assume that we compute the Fermi-arc states for the half-space above the impurity plane, i.e., for $y \geqslant 0$. The calculation for $y<0$ is not needed, since the integral is symmetric with respect to changing $y \to -y$. In order to perform the integration above, we analyze the roots of the denominator in the complex plane, as a function of $k_x, k_z$ and  $m$:
\begin{align}
z_\pm = f \pm \sqrt{f^2-1}
\end{align}
It is easy to show that 
\begin{align*}
z_+ \in (-1,\,0), z_- < -1 \quad\text{when}\; g \in (-1,\, 0), \\
z_- \in (0,\,+1), z_+ > +1 \quad\text{when}\; g \in (0,\, +1).
\end{align*}
Therefore, for the integral above we get:
\begin{align}
** = \frac{1}{g}
	\begin{cases}
		\frac{z_+^{|y|}}{z_+-z_-} &\text{for}\; g \in (-1,\, 0) \\
		\frac{z_-^{|y|}}{z_--z_+} &\text{for}\; g \in (0,\, +1)	
	\end{cases}
\end{align}

\end{document}